\documentclass[prd,preprint,superscriptaddress,showpacs,byrevtex]{revtex4}
\usepackage{bm}
\def\fsl#1{\setbox0=\hbox{$#1$}           
   \dimen0=\wd0                                 
   \setbox1=\hbox{/} \dimen1=\wd1               
   \ifdim\dimen0>\dimen1                        
      \rlap{\hbox to \dimen0{\hfil/\hfil}}      
      #1                                        
   \else                                        
      \rlap{\hbox to \dimen1{\hfil$#1$\hfil}}   
      /                                         
   \fi}                                         %
\newcommand{\be}{\begin{equation}}
\newcommand{\ee}{\end{equation}}
\newcommand{\bea}{\begin{eqnarray}}
\newcommand{\eea}{\end{eqnarray}}
\newcommand{\beq}{\begin{equation}}
\newcommand{\eeq}{\end{equation}}
\newcommand{\beqs}{\begin{eqnarray}}
\newcommand{\eeqs}{\end{eqnarray}}

\newcommand{\gsim}{\mathrel{\raisebox{-
.6ex}{$\stackrel{\textstyle>}{\sim}$}}}

\begin{document}
\title{ Proof of Factorization of $J/\Psi$ Production in Non-Equilibrium QCD at RHIC and LHC }
\author{Gouranga C Nayak }\email{nayak@max2.physics.sunysb.edu}
\affiliation{ 22 West Fourth Street \#1, Lewistown, Pennsylvania 17044, USA }
\date{\today}
\begin{abstract}
$J/\psi$ suppression/production is one of the main signature of quark-gluon plasma detection at RHIC and
LHC. In order to study $j/\psi$ suppression/production in high energy heavy-ion collisions at RHIC and LHC,
one needs to prove the factorization theorem of $j/\psi$ production in non-equilibrium QCD medium,
otherwise one will predict infinite cross section of $j/\psi$.
In this paper we prove factorization theorem of $j/\psi$ production in non-equilibrium QCD at RHIC and LHC
at all order in coupling constant.
\end{abstract}
\pacs{ 12.38.Lg; 12.38.Aw; 14.40.Pq; 12.39.St }
\maketitle
\pagestyle{plain}
\pagenumbering{arabic}
\section{Introduction}
Just after $\sim 10^{-12}$ seconds of the big bang, the universe was filled with a state of matter known as quark-gluon plasma.
The quark-gluon plasma is much hotter and denser than the ordinary matter we see today. The temperature of
the quark-gluon plasma is $\gsim $ 200 MeV ($\gsim 3.2 \times 10^{12}$ K) which is about million times larger than
the temperature of the sun ($\sim 1.56 \times 10^{7}$ K). The quark-gluon
plasma is much denser than neutron stars, besides black holes, there's nothing denser than this.
Hence it is important to recreate this early universe scenario in the laboratory,
{\it i. e.}, to produce quark-gluon plasma in the laboratory. At present RHIC
(relativistic heavy-ion colliders) at BNL and LHC (large hadron collider)
at CERN are the best facilities to produce quark-gluon plasma in the laboratory \cite{qgp,qgp1}.

RHIC and LHC use high energy heavy-ion collisions to produce quark-gluon plasma. RHIC collides two gold nuclei
at $\sqrt{s}$ = 200 GeV per nucleon \cite{phenix,star} with total energy $\sim$ 40 TeV. The LHC (in its first run)
collides two lead nuclei at $\sqrt{s}$ = 2.76 TeV per nucleon \cite{atlas,cms,alice} with total energy $\sim$ 574 TeV.
Since these huge total energies are deposited in very small volume at the initial moment of the nuclear collisions
at RHIC and LHC, the energy density necessary to create temperature $\sim $ 200 MeV to produce quark-gluon
plasma might have been achieved at RHIC and LHC. The LHC (in its second run) will collide two lead nuclei at
$\sqrt{s}$ = 5.5 TeV per nucleon with total energy $\sim$ 1150 TeV which will be even higher.

However, since the two nuclei at RHIC and LHC travel almost at speed of light, the quark-gluon plasma
produced at RHIC and LHC may be in non-equilibrium. This is because the partons inside the nuclei at RHIC
and LHC carry very high longitudinal momentum and very small transverse momentum just before the nuclear
collision. Since the hadronization takes place in a very short time scale (the typical time scale for
hadronization in QCD is $\sim 10^{-24}$ seconds), there may not be enough time for these highly
non-isotropic partons to thermalize at RHIC and LHC. Hence, in order to make meaningful comparison
of the theory with the experimental data on hadron production it may be necessary to study
nonequilibrium-nonperturbative QCD at RHIC and LHC. This, however, is a difficult problem.

Since quarks and gluons are not directly observed we can not directly detect the quark-gluon plasma at RHIC and LHC.
Hence indirect signatures are proposed for the detection of quark-gluon plasma at RHIC and LHC. The main signatures
of quark-gluon plasma detection are 1) $j/\psi$ suppression/production 2) dilepton production 3) direct photon production,
4) strangeness enhancement and 5) jet quenching etc.. In this paper we will focus on the $j/\psi$ suppression/production
signature for the quark-gluon plasma detection at RHIC and LHC.

Note that the $j/\psi$ suppression was predicted to be a signature of quark-gluon plasma detection
by using lattice QCD calculation at finite temperature in equilibrium \cite{satz}. It was argued that at high temperature the Debye
screening length is much smaller than the $j/\psi$ radius, leading to complete suppression of $j/\psi$ production
in quark-gluon plasma. However, this prediction uses lattice QCD calculation at finite temperature which may not be
applicable at RHIC and LHC where the quark-gluon plasma may be in non-equilibrium. Hence, in order to make comparison
of the theory with the experimental data on $j/\psi$ production at RHIC and LHC, it may be necessary to study $j/\psi$
production in non-equilibrium QCD.

The $j/\psi$ suppression is experimentally observed in Au-Au collisions
at $\sqrt{s}$ = 200 GeV at RHIC \cite{phenixj,starj} and in Pb-Pb collisions
at $\sqrt{s}$ = 2.76 TeV at LHC (in its first run) \cite{atlasj,cmsj,alicej}.
It is interesting to note that in most central heavy-ion collisions
the PHENIX collaboration \cite{phenixj} at RHIC has measured more $j/\psi$ suppression than the ALICE collaboration \cite{alicej}
at LHC (in its
first run). In order to explain this it is argued that there is $j/\psi$ enhancement \cite{je} and charm recombination \cite{sc}
in QCD medium etc., however these approaches need to incorporate non-perturbative QCD mechanism for
heavy quarkonium production from $c{\bar c}$ pair.
Hence the determination of in-medium properties of heavy quarkonium at RHIC and LHC
remains a challenging theoretical task.

The experimental data at RHIC and LHC suggest that the $j/\psi$
production cross section is modified in the heavy-ion collisions in comparison
to the p-p collisions at the same center of mass energy.
This modification can be due to the cold nuclear
matter effects such as nuclear shadowing, saturation, energy loss and quark-antiquark breakup
etc. and due to the hot QCD medium effects such as color
screening and secondary charm production etc.. The cold nuclear matter effects are
studied both theoretically and experimentally, for example by using
p-A collisions at the corresponding center of mass energy. The hot QCD medium effects such as
color screening is well studied theoretically by using lattice QCD at finite temperature \cite{satz,satz1},
but needs to be extended to non-equilibrium QCD to be applicable at RHIC and
LHC. Similarly secondary charm production is well studied at finite temperature \cite{je,gore,sc,sve},
but how these secondary charm-anticharm produces $j/\psi$ by using non-perturbative QCD mechanism is not well understood,
which also needs to be extended to non-equilibrium QCD at RHIC and LHC.

Hence the $j/\psi$ production mechanism
in QCD in vacuum (for example in p-p collisions) need to be modified to include QCD medium effects at
the heavy-ion collisions at RHIC and LHC.
As mentioned above, since the two nuclei at RHIC and LHC travel almost at speed of light, the quark-gluon plasma
at RHIC and LHC may be in non-equilibrium. Hence, in order to detect quark-gluon plasma by using $j/\psi$ as a
signature, one needs to study $j/\psi$ production in non-equilibrium QCD at RHIC and LHC.

Note that in order to study $j/\psi$ production in p-p collisions at high energy one needs to
prove factorization theorem. The need to prove factorization theorem arises because in the absence
of the proof of factorization theorem one will predict infinite cross section of $j/\psi$ production
\cite{collinssterman,sterman,cs2,bodwin,tucci,nayakqed,nayaksterman,nayaksterman1,nayakall}.
This is due to the interaction between charm (anticharm) quark with the nearby light quark or gluon.
The soft gluon exchange between charm (anticharm) quark and the nearby light quark (or gluon) gives
infrared divergence which makes the partonic level cross section infinite. Hence it is important to
prove that the effect of such infrared (soft) gluon exchanges either cancel or the non-canceling infrared
divergences be absorbed into the definition of the non-perturbative matrix element of the $j/\psi$ because
(soft) infrared regime corresponds to non-perturbative QCD.

In the non-equilibrium QCD medium at RHIC and LHC there are many more nearby light quarks and gluons
(in comparison to the corresponding situation in p-p collisions in QCD in vacuum)
with which the charm (anticharm) quark interacts. Because of this,
infrared divergences seem to be severe at RHIC and LHC heavy-ion collisions than that
in corresponding p-p collisions. Hence it is necessary
to prove that the infrared divergences due to the soft gluons
exchange between charm (anticharm) quark and the nearby light quark (or gluon) in non-equilibrium
QCD medium at RHIC and LHC either cancel or the non-canceling infrared divergences
be absorbed into the definition of the non-perturbative matrix element
of the $j/\psi$ production in non-equilibrium QCD.

Since the $j/\psi$ suppression as a
signature of quark-gluon plasma detection was suggested by using
$j/\psi$ from color singlet $c{\bar c}$ pair in \cite{satz}, by comparing with the correlation length
obtained from lattice QCD \cite{satz2}, we will consider $j/\psi$
production from color singlet $c{\bar c}$ pair in non-equilibrium QCD in this paper.
We will prove, in this paper, that the infrared divergences due to the soft gluons
exchange between charm quark and the nearby light-like quark (or gluon) in non-equilibrium
QCD at RHIC and LHC exactly cancel with the corresponding infrared divergences
due to the soft gluons exchange between anticharm quark and the same nearby light-like quark (or gluon)
at all order in coupling constant in the
$j/\psi$ production from color singlet $c{\bar c}$ pair in non-equilibrium QCD.
This proves the factorization theorem of $j/\psi$ production in non-equilibrium
QCD at RHIC and LHC at all order in coupling constant.

The paper is organized as follows. In section II we briefly discuss path integral formulation of non-equilibrium QCD.
In section III we discuss the non-perturbative matrix element
of $j/\psi$ production from color singlet charm-anticharm pair in non-equilibrium QCD.
In section IV we discuss how
longitudinal polarization of the gauge field in quantum field theory is used to describe infrared
divergence. In section V we discuss the gauge field generated by the eikonal current of the
light-like Wilson line in quantum field theory.
In section VI we prove factorization theorem of $j/\psi$ production at high energy colliders.
In section VII we prove factorization theorem of $j/\psi$ production in non-equilibrium QCD at RHIC
and LHC. Section VIII contains conclusions.

\section{ Non-Equilibrium QCD Using Closed-Time Path Integral Formalism }

Unlike $pp$ collisions, the ground state at RHIC and LHC heavy-ion collisions
(due to the presence of a QCD medium at initial time $t=t_{in}$ (say $t_{in}$=0)
is not a vacuum state $|0>$ any more.
We denote $|in>$ as the initial state of the non-equilibrium QCD
medium at $t_{in}$.

Consider massless scalar field theory first. In the closed-time path (CTP) formalism
the generating functional in the path integral formulation
for scalar field theory in non-equilibrium is given by
\bea
Z[\rho,J_+,J_-]=\int [d\phi_+][d\phi_-] ~{\rm exp}[i[S[\phi_+]-S[\phi_-]+\int d^4x J_+\phi_+-\int d^4x J_-\phi_-]]~<\phi_+,0|\rho|0,\phi_-> \nonumber \\
\label{rho7}
\eea
where $\rho$ is the initial density of state,
$S[\phi]$ is the full action in scaler field theory and $|\phi_{\pm},0>$ is the quantum state corresponding to the field
configuration $\phi_{\pm}(\vec{x},t=0)$. Since there are two sources $J_+$ and $J_-$ corresponding to two time branches $+$ and
$-$ in non-equilibrium quantum field theory one finds that there are four Green's functions in non-equilibrium.
The four Green's functions in non-equilibrium can be obtained from the generating functional
as follows
\bea
&& G_{++}(x,x') = \frac{\delta Z[\rho,J_+,J_-]}{i^2 \delta J_+(x) J_+(x')}=<in|T\phi (x) \phi (x')|in> \nonumber \\
&& G_{--}(x,x') = \frac{\delta Z[\rho,J_+,J_-]}{(-i)^2 \delta J_-(x) J_-(x')}= <in|{\bar T} \phi (x) \phi (x')|in> \nonumber \\
&& G_{+-}(x,x') = \frac{\delta Z[\rho,J_+,J_-]}{-i^2 \delta J_+(x) J_-(x')}= <in|\phi (x') \phi (x)|in> \nonumber \\
&& G_{-+}(x,x') = \frac{\delta Z[\rho,J_+,J_-]}{-i^2 \delta J_-(x) J_+(x')}= <in|\phi (x) \phi (x')|in>
\label{green}
\eea
where $+$($-$) sign corresponds to upper(lower) time branch of the Schwinger-Keldysh
closed-time path \cite{schw,keldysh}, $T$ is the time order product and ${\bar T}$ is the
anti-time order product.

The generating functional in non-equilibrium QCD (including heavy quark)
in the path integral formulation is given by \cite{greiner,cooper}
\bea
&& Z[\rho,J_+,J_-,\eta_+^u,\eta_-^u,{\bar \eta}_+^u,{\bar \eta}_-^u,\eta_+^d,\eta_-^d,{\bar \eta}_+^d,{\bar \eta}_-^d,\eta_+^s,\eta_-^s,{\bar \eta}_+^s,{\bar \eta}_-^s,\eta_+^h,\eta_-^h,{\bar \eta}_+^h,{\bar \eta}_-^h]\nonumber \\
&&=\int [dQ_+] [dQ_-][d{\bar \psi}_{1+}] [d{\bar \psi}_{1-}] [d \psi_{1+} ] [d\psi_{1-}][d{\bar \psi}_{2+}] [d{\bar \psi}_{2-}] [d \psi_{2+} ] [d\psi_{2-}][d{\bar \psi}_{3+}] [d{\bar \psi}_{3-}] [d \psi_{3+} ] [d\psi_{3-}]\nonumber \\
&&
[d{\bar \Psi}_{+}] [d{\bar \Psi}_{-}] [d \Psi_{+} ] [d\Psi_{-}]
\times ~{\rm det}(\frac{\delta \partial_\mu Q_+^{\mu a}}{\delta \omega_+^b})~\times ~{\rm det}(\frac{\delta \partial_\mu Q_-^{\mu a}}{\delta \omega_-^b}) \nonumber \\
&& {\rm exp}[i\int d^4x [-\frac{1}{4}({F^a}_{\mu \nu}^2[Q_+]-{F^a}_{\mu \nu}^2[Q_-])-\frac{1}{2 \alpha} (
(\partial_\mu Q_+^{\mu a })^2-(\partial_\mu Q_-^{\mu a })^2) \nonumber \\
&&+\sum_{l=1}^3{\bar \psi}_{l+}  [i\gamma^\mu \partial_\mu -m_l +gT^a\gamma^\mu Q^a_{\mu +}]  \psi_{l+} -\sum_{l=1}^3{\bar \psi}_{l-}  [i\gamma^\mu \partial_\mu -m_l +gT^a\gamma^\mu Q^a_{\mu -}]  \psi_{l-}\nonumber \\
&&+{\bar \Psi}_{+}  [i\gamma^\mu \partial_\mu -M +gT^a\gamma^\mu Q^a_{\mu +}]  \Psi_{+}-{\bar \Psi}_{-}  [i\gamma^\mu \partial_\mu -M +gT^a\gamma^\mu Q^a_{\mu -}]  \Psi_{-}
+ J_+ \cdot Q_+ -J_- \cdot Q_-\nonumber \\
&&+\sum_{l=1}^3[{\bar \eta}_{l+} \cdot \psi_{l+} - {\bar \eta}_{l-} \cdot \psi_{l-} +  {\bar \psi}_{l+} \cdot \eta_{l+}
- {\bar \psi}_{l-} \cdot \eta_{l-}]+{\bar \eta}_{h+} \cdot \Psi_{+} - {\bar \eta}_{h-} \cdot \Psi_{-} +  {\bar \Psi}_{+} \cdot \eta_{h+}
- {\bar \Psi}_{-} \cdot \eta_{h-}]]
\nonumber \\
&& \times ~<Q_+,\psi_+^u,{\bar \psi}_+^u,\psi_+^d,{\bar \psi}_+^d,\psi_+^s,{\bar \psi}_+^s,\Psi_+,{\bar \Psi}_+,0|~\rho~|0,\Psi_-,{\bar \Psi}_-,{\bar \psi}_-^s,\psi_-^s,{\bar \psi}_-^d,\psi_-^d,{\bar \psi}_-^u,\psi_-^u,Q_-> \nonumber \\
\label{zfqinon}
\eea
where $\rho$ is the initial density of state. The state $|Q^\pm,\psi^\pm,{\bar \psi}^\pm,0>$ corresponds to the field
configurations $Q_\mu^a({\vec x},t=t_{in}=0)$, $\psi({\vec x},t=t_{in}=0)$ and ${\bar \psi}({\vec x},t=t_{in}=0)$ respectively and
$J^{\mu a}(x)$ is the external source for the quantum gluon field $Q^{\mu a}(x)$ and the ${\bar \eta}^u,~{\bar \eta}^d,~{\bar \eta}^s$ are external sources for $l=1,2,3=u,d,s$ quark fields respectively and ${\bar \eta}_h$ is the external source for the heavy quark field $\Psi$, $m_l$
is the mass of the light quark of type $l$, the $M$ is the mass of the heavy quark and
\bea
F^a_{\mu \nu}[Q]=\partial_\mu Q^a_\nu(x)-\partial_\nu Q^a_\mu(x)+gf^{abc}Q^b_\mu(x)Q^c_\nu(x),~~~~~~~~~{F^a}_{\mu \nu}^2[Q]={F}^{\mu \nu a}[Q]{F}^a_{\mu \nu}[Q].
\eea
Note that we work in the frozen ghost formalism \cite{greiner,cooper} for the medium part at the initial time $t=t_{in}=0$.

For the heavy quark Dirac field $\Psi_r(x)$ in non-equilibrium QCD,
the nonequilibrium-nonperturbative matrix element of the type $<in|{\bar \Psi}_r(x_1) \Psi_r(x_1) {\bar \Psi}_s(x_2) \Psi_s(x_2)|in>$
is given by \cite{tucci}
\bea
&&<in|{\bar \Psi}_r(x_1) \Psi_r(x_1) {\bar \Psi}_s(x_2) \Psi_s(x_2)|in>\nonumber \\
&&=\frac{\delta}{\delta \eta_{hr}(x_1)}\frac{\delta}{\delta {\bar \eta}_{hr}(x_1)}\frac{\delta}{\delta \eta_{hs}(x_2)}\frac{\delta}{\delta {\bar \eta}_{hs}(x_2)}
Z[\rho,J_+,J_-,\eta_+^u,\eta_-^u,{\bar \eta}_+^u,{\bar \eta}_-^u,\eta_+^d,\eta_-^d,{\bar \eta}_+^d,{\bar \eta}_-^d,\eta_+^s,\eta_-^s,{\bar \eta}_+^s,{\bar \eta}_-^s\nonumber \\
&&,\eta_+^h,\eta_-^h,{\bar \eta}_+^h,{\bar \eta}_-^h]|_{J_+=J_-=\eta_+^u=\eta_-^u={\bar \eta}_+^u={\bar \eta}_-^u=\eta_+^d=\eta_-^d={\bar \eta}_+^d={\bar \eta}_-^d=\eta_+^s=\eta_-^s={\bar \eta}_+^s={\bar \eta}_-^s=\eta_+^h=\eta_-^h={\bar \eta}_+^h={\bar \eta}_-^h=0}
\label{malepg}
\eea
where the repeated (closed-time path) indices $r,s=+,-$ are not summed
and the suppression of the normalization factor $Z[0]$ is understood as it will cancel in the final result, see eq. (\ref{finalz}).

\section{ Non-Perturbative Matrix Element of $J/\Psi$ Production From Color Singlet Charm-Anticharm Pair in Non-Equilibrium QCD}

The charm-anticharm quark ($c{\bar c}$) pair production in the initial collision of the two nuclei at RHIC and LHC
can be calculated by using pQCD. The spectral notation of $j/\psi$ is given by $^{2S+1}L_J$ where spin quantum number
$S=1$ (spin triplet), orbital angular momentum quantum number $L=0$ ($S$-wave) and total angular momentum quantum number $J=1$.
Once the $c{\bar c}$ pair is formed in the color singlet and spin triplet state
it can form $j/\psi$. The formation of $j/\psi$ from $c{\bar c}$ involves non-perturbative QCD mechanism
which can not be calculated by using pQCD.

Note that the potential model calculations \cite{singf} use phenomenological potentials like Coulomb potential or Coulomg plus linear
potential etc. between color singlet $c{\bar c}$ by solving non-relativistic wave equation to determine the radial
wave function at the origin $R(0)$ for the $j/\psi$ formation from color singlet $c{\bar c}$ pair. However, strictly speaking, unlike
QED, the exact form of the potential energy between $c{\bar c}$ pair is not known in QCD because we do not know
the exact form of the classical Yang-Mills potential $A^{\mu a}(x)$ yet, although Yang-Mills theory was discovered
almost 60 years ago. Hence, unlike QED, we depend on the experimental data in QCD \cite{singf,pot}
to determine the non-perturbative matrix
element of the heavy quarkonium production from heavy quark and antiquark pair. The non-perturbative matrix
element for $j/\psi$ formation from color singlet $c{\bar c}$ pair needs to be universal, {\it i. e.}, it should not
change from one experiment to other. Hence it is necessary to use the correct definition of the non-perturbative matrix
element of $j/\psi$ production from color singlet and spin triplet $c{\bar c}$ pair. If one does not use the
correct definition of the non-perturbative matrix
element of $j/\psi$ production from color singlet and spin triplet $c{\bar c}$ pair then one will predict infinite
cross section for $j/\psi$ production. This can be seen as follows.

If the factorization theorem holds, then the production cross section for $j/\psi$ at transverse momentum $P_T$ can be
written as the product of perturbative cross section of $c{\bar c}$ production times the universal non-perturbative matrix element,
\bea
d\sigma_{A+B \rightarrow j/\psi +X(P_T)} = d{\hat \sigma}_{A+B \rightarrow c{\bar c}[^{3}S_1] +X(P_T)} ~<0|{\cal O}_{j/\psi}|0>
\label{css}
\eea
where $d{\hat \sigma}_{A+B \rightarrow c{\bar c}[^{3}S_1] +X(P_T)}$ is the perturbative cross section for $c{\bar c}$
pair production in color singlet and spin triplet state and the
non-perturbative matrix element $<0|{\cal O}_{j/\psi}|0>$ represents the probability of $c{\bar c}$ pair in
color singlet and spin triplet state to produce $j/\psi$. In eq. (\ref{css}) the notation $^3S_1$ stands for
the spectral notation $^{2S+1}L_J$ of the $j/\psi$ where $S=1,L=0,J=1$ are the spin, orbital angular momentum and total
angular momentum respectively.

The definition of the non-perturbative matrix element $<0|{\cal O}_{j/\psi}|0>$ can be found as follows. The amplitude
for the $c{\bar c}$ to form the specific hadron $H$ plus other hadrons X is given by
\bea
<H+X|c{\bar c}>.
\label{amp}
\eea
The amplitude in eq. (\ref{amp}) gives the probability of the hadron $H$ production from the $c{\bar c}$ by using the
non-perturbative matrix element
\bea
&&<0|{\cal O}_{H}|0> = \sum_X <c{\bar c}|H+X><H+X|c{\bar c}>=\sum_X <c{\bar c}|a^\dagger_H|X><X|a_H|c{\bar c}>\nonumber \\
&&=<c{\bar c}|a^\dagger_Ha_H|c{\bar c}>
\label{lll}
\eea
where $a^\dagger_H$ is the creation operator of the hadron.
Writing the Dirac wave function of the quark in terms of creation and annihilation operators $a^\dagger,~a$
and then using the definition of the charm quark state $|c>$ formed from the QCD vacuum $|0>$ by using the creation operator
$a^\dagger$:
\bea
|c> =a^\dagger |0>
\label{inout0}
\eea
we find from eq. (\ref{lll}) that the definition of the non-perturbative matrix element at the origin
for the $\eta_c$ production from color singlet $c{\bar c}$ pair is given by
\bea
<0|{\cal O}_{H}|0> =<0|\chi^\dagger \psi a^\dagger_Ha_H \psi^\dagger \chi |0>
\label{allle}
\eea
where $\psi$, $\chi$ are two component Dirac spinors of the heavy quark wave function.
Similar to the derivation of eq. (\ref{allle}) we find that the
definition of the non-perturbative matrix element at the origin
for the $j/\psi$ production from $c{\bar c}$ pair in color singlet and spin triplet state is given by
\bea
<0|{\cal O}_{H}|0> =<0|\chi^\dagger \sigma^i \psi a^\dagger_Ha_H \psi^\dagger \sigma^i \chi |0>
\label{alll}
\eea
where $\sigma^i$ is the Pauli spin matrix.

The wave function at the origin $R(0)$ for the $j/\psi$ production from $c{\bar c}$
pair in color singlet mechanism is given by
\bea
|R(0)|^2 = \frac{2\pi }{N_c} <0|\chi^\dagger \sigma^i \psi a^\dagger_Ha_H \psi^\dagger \sigma^i \chi |0>.
\label{wf}
\eea

Note that in the perturbative cross section $d{\hat \sigma}_{A+B \rightarrow c{\bar c}[^{3}S_1]+X(P_T)} $ in eq. (\ref{css})
the infrared divergences can occur due to the interaction of the $c$ and/or ${\bar c}$ with nearby light quark or gluon.
This infrared divergences occur due to the soft (infrared) gluon interaction between the $c$ and/or ${\bar c}$ with nearby light quark or gluon.
Hence it is necessary to prove that any non-canceling infrared divergences in the perturbative cross
section $d{\hat \sigma}_{A+B \rightarrow c{\bar c}[^{3}S_1]+X(P_T)} $ in eq. (\ref{css}) cancels with the corresponding infrared divergences
in the non-perturbative matrix element $<0|{\cal O}_{j/\psi}|0>$ in eq. (\ref{css}). If such a cancelation does not happen then
one will predict infinite cross section $d\sigma_{A+B \rightarrow j/\psi +X(P_T)}$
for $j/\psi$ production from eq. (\ref{css}). Such cancelation of infrared divergences is called factorization.
Hence it is necessary to prove factorization theorem of $j/\psi$ production at high energy colliders.

Similar to eq. (\ref{inout0}) in vacuum, using the definition of the $|c>$ state in non-equilibrium
by using the creation operator $a^\dagger$:
\bea
|c> =a^\dagger |in>
\label{inoutm}
\eea
we find from eq. (\ref{lll}) that the non-perturbative matrix element for $\eta_c$ production from color singlet $c{\bar c}$ pair
in non-equilibrium QCD is given by
\bea
<in|{\cal O}_{H}|in> =<in|\chi^\dagger  \psi a^\dagger_Ha_H \psi^\dagger  \chi |in>
\label{mallle}
\eea
where $|in>$ is the initial state of the non-equilibrium QCD medium. Eq. (\ref{mallle}) is similar to
eq. (\ref{allle}) except that the vacuum expectation is replaced by medium average. Similar to the
derivation of eq. (\ref{mallle}) we find that
the non-perturbative matrix element for $j/\psi$ production from $c{\bar c}$ pair in color singlet
and spin triplet state in non-equilibrium QCD is given by
\bea
<in|{\cal O}_{H}|in> =<in|\chi^\dagger \sigma^i \psi a^\dagger_Ha_H \psi^\dagger \sigma^i \chi |in>
\label{malll}
\eea
which is similar to eq. (\ref{alll}) except that the vacuum expectation is replaced by medium average.

The infrared divergences can occur due to the (soft) gluon exchange between the $c$ and/or ${\bar c}$ with nearby light quark or gluon
in non-equilibrium QCD medium at RHIC and LHC.
Hence it is necessary to prove that any non-canceling infrared divergences due to these soft gluons exchange
in non-equilibrium QCD medium cancel with the corresponding infrared divergences
in the non-perturbative matrix element $<in|{\cal O}_{j/\psi}|in>$. If such a cancelation does not happen then
one will predict infinite cross section for $j/\psi$ production at RHIC and LHC.
Such cancelation of infrared divergences is called factorization in non-equilibrium QCD.
Hence it is necessary to prove factorization theorem of $j/\psi$ production in non-equilibrium QCD at RHIC and LHC.

In this paper we will prove that the infrared divergences due to the soft gluons
exchange between charm quark and the nearby light-like quark (or gluon) in non-equilibrium
QCD at RHIC and LHC exactly cancel with the corresponding infrared divergences
due to the soft gluons exchange between anticharm quark and the same nearby light-like quark (or gluon)
at all order in coupling constant in the
$j/\psi$ production from color singlet $c{\bar c}$ pair in non-equilibrium QCD.
This proves the factorization theorem of $j/\psi$ production in non-equilibrium
QCD at RHIC and LHC at all order in coupling constant.

\section{ Infrared Divergence and Unphysical Longitudinal Polarization of the Gauge Field in Quantum Field Theory }

Proof of factorization theorem is usually given in diagrammatic method by using
pQCD \cite{collinssterman,sterman,cs2,bodwin,nayaksterman}. However,
since the matrix element $<0|\chi^\dagger \sigma^i \psi a^\dagger_Ha_H \psi^\dagger \sigma^i \chi |0>$
for $j/\psi$ production from $c{\bar c}$ pair in eq. (\ref{alll}) is the non-perturbative matrix element it is useful
to use the path integral formulation of QCD to prove factorization theorem. In fact, the proof of factorization theorem using path
integral formulation is enormously simplified \cite{nayakall} in comparison to diagrammatic method. Also
path integral method naturally proves factorization theorem at all order in coupling constant whereas the
diagrammatic method using pQCD suffers difficulties beyond certain order of coupling constant calculation \cite{nayaksterman}.

Below we will briefly describe the general technique used in the path integral formulation in
any quantum field theory to prove factorization theorem.
Since we are interested in the soft-gluons exchange between the charm (and/or anticharm) quark with the
nearby light-like quark and/or gluon
we need to study the infrared behavior of the non-perturbative correlation
function of the type $<0|{\bar \Psi}(x_1)\Psi(x_1){\bar \Psi}(x_2)\Psi(x_2)|0>$ in QCD in the presence of
light-like Wilson line, in order to study $j/\psi$ production from color singlet $c{\bar c}$ pair.
Since a light-like charge produces pure gauge field in classical mechanics \cite{collinssterman,nayakj,nayake} and
in quantum field theory (see the next section), one can study the infrared behavior of the non-perturbative correlation
function of the type $<0|{\bar \Psi}(x_1)\Psi(x_1){\bar \Psi}(x_2)\Psi(x_2)|0>$ in QCD by using the path integral
formulation of the background field method of QCD in the presence of pure gauge background field in order to
prove factorization infrared divergences at all order in coupling constant \cite{nayakall}.

Note that the pure gauge field in quantum field theory corresponds to longitudinal polarization of the
gauge field. In this section we will show how the longitudinal polarization of the gauge field is used in quantum field theory to describe
soft (infrared) divergences. Since the argument is valid in any quantum field theory let us see how the pure gauge field
is used in QED to prove factorization of infrared divergences in QED. Consider an incoming electron of four
momentum $p^\mu$ and mass $m$ emitting a real photon of four momentum $k^\mu$ in QED.
The corresponding Feynman diagram contribution is given by \cite{grammer}
\begin{eqnarray}
&& {\cal M}=\frac{1}{\gamma_\nu p^\nu -\gamma_\nu k^\nu -m} \gamma_\mu \epsilon^\mu(k)u(p)=-\frac{p \cdot \epsilon(k)}{p \cdot k}u(p)+\frac{k^\nu \gamma_\nu \gamma_\mu \epsilon^\mu(k)}{2p \cdot k}u(p)
\label{total}
\end{eqnarray}
where we write
\begin{eqnarray}
{\cal M}_{\rm eikonal}=-\frac{p \cdot \epsilon(k)}{p \cdot k}u(p)
\label{eik}
\end{eqnarray}
and
\begin{eqnarray}
{\cal M}_{\rm non-eikonal}=\frac{k^\nu \gamma_\nu \gamma_\mu \epsilon^\mu(k)}{2p \cdot k} u(p).
\label{noneik}
\end{eqnarray}
From eq. (3.2) of \cite{grammer} we write the gauge field as
\begin{equation}
\epsilon^\mu(k) = [\epsilon^\mu(k) -k^\mu \frac{p \cdot \epsilon(k)}{p \cdot k}]+k^\mu \frac{p \cdot \epsilon(k)}{p \cdot k}=\epsilon_{\rm phys}^\mu(k)+\epsilon_{\rm pure}^\mu(k)
\label{1}
\end{equation}
where
\begin{equation}
\epsilon_{\rm phys}^\mu(k) = [\epsilon^\mu(k) -k^\mu \frac{p \cdot \epsilon(k)}{p \cdot k}]
\label{1a}
\end{equation}
is the physical gauge field [corresponding to transverse polarization of the gauge field] and
\begin{equation}
\epsilon_{\rm pure}^\mu(k) = k^\mu \frac{p \cdot \epsilon(k)}{p \cdot k}
\label{1b}
\end{equation}
is the pure gauge field [corresponding to longitudinal polarization of the gauge field].

Now using eq. (\ref{1}) in (\ref{total}) we find that the total contribution of the Feynman diagram is given by
\begin{eqnarray}
&& {\cal M}= {\cal M}_{\rm eikonal}+{\cal M}_{\rm non-eikonal}
\label{totala}
\end{eqnarray}
where
\begin{eqnarray}
{\cal M}_{\rm eikonal} = -\frac{p \cdot \epsilon_{\rm phys}(k)}{p \cdot k}u(p) - \frac{p \cdot \epsilon_{\rm pure}(k)}{p \cdot k}u(p)=-\frac{p \cdot \epsilon_{\rm pure}(k)}{p \cdot k}u(p)
\label{totalb}
\end{eqnarray}
and
\begin{eqnarray}
{\cal M}_{\rm non-eikonal} = \frac{k^\nu \gamma_\nu \gamma_\mu \epsilon_{\rm phys}^\mu(k)}{2p \cdot k} u(p)+\frac{k^\nu \gamma_\nu \gamma_\mu \epsilon_{\rm pure}^\mu(k)}{2p \cdot k} u(p)=\frac{k^\nu \gamma_\nu \gamma_\mu \epsilon_{\rm phys}^\mu(k)}{2p \cdot k} u(p).
\label{totalc}
\end{eqnarray}

Hence in the soft photon limit $(k_0,k_1,k_2,k_3) \rightarrow 0$ we find from the eqs. (\ref{total}) and (\ref{totalb}) that
\begin{eqnarray}
-{\cal M}_{\rm eikonal}=\frac{p \cdot \epsilon(k)}{p \cdot k}u(p) =\frac{p \cdot \epsilon_{\rm pure}(k)}{p \cdot k}u(p) \rightarrow \infty ~~~~~~~~~~~{\rm as}~~~~~~~~~~(k_0,k_1,k_2,k_3) \rightarrow 0
\label{totald}
\end{eqnarray}
which implies that the physical gauge field [corresponding to transverse polarization] does not contribute to
the soft (infrared) divergences in quantum field theory and the soft (infrared) divergences can be calculated
by using pure gauge field [corresponding to longitudinal polarization] in quantum field theory.

Similarly from eqs. (\ref{total}) and (\ref{totalc}) we find that
\begin{eqnarray}
{\cal M}_{\rm non-eikonal}=\frac{k^\nu \gamma_\nu \gamma_\mu \epsilon^\mu(k)}{2p \cdot k} u(p)=\frac{k^\nu \gamma_\nu \gamma_\mu \epsilon_{\rm phys}^\mu(k)}{2p \cdot k} u(p) \rightarrow {\rm finite} ~~~~~~~~~~~{\rm as}~~~~~~~~~~(k_0,k_1,k_2,k_3) \rightarrow 0\nonumber \\
\label{totale}
\end{eqnarray}
which contribute to the finite part of the cross section which implies that pure gauge field [corresponding to longitudinal
polarization] does not contribute to the finite cross section and the finite cross section can be calculated by using
physical gauge field [corresponding to transverse polarization].

Hence we find that the non-eikonal-line part of the diagram as given by eq. (\ref{totale}) is necessary if we are calculating
the finite value of the cross section but it is not necessary if we are calculating the relevant infrared divergence behavior. The relevant
infrared divergence behavior can be calculated by using the eikonal-line part of the diagram as given by eq. (\ref{totald}).

Hence we find that we do not need to calculate the finite value of the cross section
(or the full cross section) [which will require the non-eikonal-line part of the diagram as given
by eq. (\ref{totale})] to study the relevant infrared divergence behavior. The relevant infrared
divergence behavior can be calculated by using eikonal approximation as given by eq. (\ref{totald}).

From eq. (\ref{totale}) we find that
\begin{eqnarray}
{\cal M}^{\rm pure~gauge~field}_{\rm non-eikonal}=\frac{k^\nu \gamma_\nu \gamma_\mu \epsilon^\mu_{\rm pure}(k)}{2p \cdot k} u(p)=0.
\label{totalf}
\end{eqnarray}
We are interested in the infrared divergence behavior due to the presence of the light-like Wilson line.
We will show in the next section that the eikonal current of the light-like charge generates pure gauge
field in quantum field theory. Hence from eqs. (\ref{totald}) and (\ref{totalf}) we find that the
soft (infrared) divergence behavior due to the presence light-like Wilson line
can be studied by using pure gauge field in quantum field theory without modifying the finite value of the cross section

\section{ Gauge Field Generated by Eikonal Current of Light-Like Wilson Line in Quantum Field Theory }

In order to study factorization of infrared divergences by using the background field method of QED,
the soft photon cloud traversed by the electron is represented by the pure gauge background field
\cite{tucci} due to the presence of the light-like Wilson line.
As mentioned above, in classical mechanics the assertion that the gauge field that is produced
by a highly relativistic (light-like) charge is a pure gauge field at all time-space position
$x^\mu$ except at the position transverse to the motion of the charge
at the time of closest approach \cite{collinssterman,nayakj,nayake}.
One may ask a question if this assertion is correct in quantum field theory. In this section
we will show that this assertion is correct in quantum field theory. We will use path integral formulation
of the quantum field theory to show this.

The generating functional for the gauge field
in the quantum field theory in the presence of external source $J^\mu(x)$ in the path integral formulation is given by
\bea
Z[J]=\int [dA]
e^{i\int d^4x [-\frac{1}{4}{F}_{\mu \nu}^2[A] -\frac{1}{2 \alpha} (\partial_\mu A^{\mu })^2+ J \cdot A ]}
\label{zfqv}
\eea
where
\bea
F^{\mu \nu}[A]=\partial^\mu A^\nu(x)-\partial^\nu A^\mu(x),~~~~~~~~~{F}_{\mu \nu}^2[A]={F}^{\mu \nu}[A]{F}_{\mu \nu}[A].
\eea
The effective action $S_{eff}[J]$ is given by \cite{peter}
\bea
<0|0>_J =\frac{Z[J]}{Z[0]}=e^{iS_{eff}[J]}
\label{vacv}
\eea
where
\bea
S_{eff}[J] = -\frac{1}{2} \int d^4x d^4x' J^\mu(x) D_{\mu \nu}(x-x')J^\nu(x')
\label{wja}
\eea
where $D_{\mu \nu}(x-x')$ is the photon propagator.

The photon propagator in the coordinate space is given by
\bea
D_{\mu \nu}(x-x') = \frac{1}{\partial^2}[g_{\mu \nu}+\frac{(\alpha -1)}{\partial^2}\partial_\mu \partial_\nu] \delta^{(4)}(x-x').
\label{dmnc}
\eea
Using eq. (\ref{dmnc}) in (\ref{wja}) we find
\bea
S_{eff}[J] = -\frac{1}{2} \int d^4x J^\mu(x)\frac{1}{\partial^2} [g_{\mu \nu}+\frac{(\alpha -1)}{\partial^2}\partial_\mu \partial_\nu] J^\nu(x).
\label{wjh}
\eea
From the continuity equation we have
\bea
\partial_\mu J^\mu(x)=0.
\label{ceq}
\eea
Using eq. (\ref{ceq}) in (\ref{wjh}) we find
\bea
S_{eff}[J] = -\frac{1}{2} \int d^4x J^\mu(x)  \frac{1}{\partial^2}J_\mu(x).
\label{wj}
\eea

First of all, by using the path integral formulation of the quantum field theory we will derive Coulomb's
law for static charge. Note that the derivation of the Coulomb's law by using path integral
formulation of the quantum field theory is not necessary to prove factorization theorem.
We have included it here only to demonstrate the correctness of the prediction of the
path integral formulation in quantum field theory which we will use (see below)
to show that the eikonal current of the light-like charge generates pure gauge field
in quantum field theory.

In order to derive Coulomb's law by using path integral formulation
of the quantum field theory let us consider the static charge at the position
${\vec X}$. The current density for this static charge is given by
\bea
J^\mu(x) =e\delta^{\mu 0} \delta^{(3)}({\vec x}-{\vec X}).
\label{cdr}
\eea
Using eq. (\ref{cdr}) in (\ref{wj}) we find
\bea
{S}_{eff}[J] = \frac{e^2}{2}\int d^4x \delta^{(3)}({\vec x}-{\vec X})
 \frac{1}{\nabla^2}\delta^{(3)}({\vec x}-{\vec X})
 =- \frac{e^2}{2}\int d^4x [\nabla \frac{1}{|{\vec x}-{\vec X}|}] \cdot \nabla \frac{1}{|{\vec x}-{\vec X}|}\nonumber \\
\eea
which gives the effective lagrangian density
\bea
{\cal L}_{eff}(x)
 =- \frac{e^2}{2} [\nabla \frac{1}{|{\vec x}-{\vec X}|}] \cdot \nabla \frac{1}{|{\vec x}-{\vec X}|} =-\frac{E^2(x)}{2}
\eea
which reproduces the Coulomb's law. Hence we have shown that the assertion that a charge
at rest generates a Coulomb gauge field is correct in quantum field theory.

Similarly using the above procedure in quantum field theory we will show that the assertion that the
eikonal current of the light-like charge generates pure gauge field is correct in quantum field theory. 
This can be shown as follows.

In QED the infrared (or soft) divergence arises only from the emission of a photon for which all components of the four-momentum
are small. As described in the previous section, the Eikonal propagator times the Eikonal vertex for a soft photon with
momentum $k$ interacting with a light-like electron moving with four momentum $p^\mu$ is given by
\cite{collins,tucci,collinssterman,berger,bodwin,grammer,frederix,nayakqed,scet1,pathorder,nayaka2,nayaka3}
\bea
e~\frac{p^\mu}{p \cdot k+i\epsilon }=e~\frac{l^\mu}{l \cdot k+ i\epsilon }
\label{eikonaliq}
\eea
where $l^\mu$ is the four-velocity of the light-like electron. Note that when we say the "light-like electron" we mean the electron
that is traveling at its highest speed which is arbitrarily close to the speed of light
($|{\vec l}|\sim 1$) as it can not travel exactly at speed of light ($|{\vec l}|= 1$)
because it has finite mass even if the mass of the electron is very small. From eq. (\ref{eikonaliq})
we find
\bea
&& e\int \frac{d^4k}{(2\pi)^4} \frac{l\cdot {  A}(k)}{l\cdot k +i\epsilon } =-e i\int_0^{\infty} d\lambda \int \frac{d^4k}{(2\pi)^4} e^{i l \cdot k \lambda} l\cdot {A}(k) = ie\int_0^{\infty} d\lambda l\cdot { A}(l\lambda)
\label{ftgtq}
\eea
where the photon field  $ { A}^{\mu }(x)$ and its Fourier transform $ { A}^{\mu }(k)$ are related by
\bea
{ A}^{\mu }(x) =\int \frac{d^4k}{(2\pi)^4} { A}^{\mu }(k) e^{ik \cdot x}.
\label{ftq}
\eea
From eq. (\ref{ftgtq}) we find
\bea
ie\int_0^{\infty} d\lambda l\cdot { A}(l\lambda)=i \int d^4x J^\mu(x) A_\mu(x)
\label{ftgtqmn}
\eea
where the eikonal current density $J^\mu(x)$ of the light-like charge $e$ is given by
\bea
J^\mu(x) = el^\mu \int d\lambda ~\delta^{(4)}(x-l\lambda), ~~~~~~~~~~~~~l^2=l^\mu l_\mu=0.
\label{ekcd}
\eea

Hence by using the path integral formulation of the quantum field theory we find by using eq.
(\ref{ekcd}) in (\ref{wj}) that the effective action is given by
\bea
&&{S}_{eff}[J] = -l^2\frac{e^2}{2}\int d^4x \int d\lambda ~\delta^{(4)}( x-l\lambda)
\frac{1}{\partial^2} \int d\lambda' ~\delta^{(4)}({ x}-l\lambda') \nonumber \\
&&=l^2\frac{e^2}{2}\int d^4x [\partial^\mu \frac{1}{l \cdot (x-l\frac{x^2}{2l \cdot x})}] [\partial_\mu \frac{1}{l \cdot (x-l\frac{x^2}{2l \cdot x})}]
\eea
which gives the effective lagrangian density
\bea
{\cal L}_{eff}(x)= l^2\frac{e^2}{2} \frac{[l^\mu -\frac{l^2}{l \cdot x}(x^\mu -l^\mu\frac{x^2}{2l \cdot x})][l_\mu -\frac{l^2}{l \cdot x}(x_\mu -l_\mu\frac{x^2}{2l \cdot x})]}{[l \cdot x-l^2\frac{x^2}{2l \cdot x}]^4}.
\label{eeffl}
\eea
From eq. (\ref{eeffl}) we find the effective lagrangian density
\bea
{\cal L}_{eff}(x) = \frac{e^2}{2}\frac{(l^2)^2}{(l \cdot x)^4},~~~~~~~~~~~~~~~~{\rm for}~~~~~~~~~~~~~~~l\cdot x \neq 0
\label{wjlk}
\eea
which gives
\bea
{\cal L}_{eff}(x) = 0
\label{wjl}
\eea
at all time-space position $x^\mu$ except at the position transverse to the motion of the charge (${\vec l} \cdot {\vec x}=0$)
at the time of closest approach ($x_0=0$) where we have used
\bea
l^2=l^\mu l_\mu=0
\eea
where $l^\mu=\frac{v^\mu_c}{c}$ with $v^\mu_c=(c,{\vec v}_c)$ where ${\vec v}_c^2=c^2$ and $c$ is the speed of light.
In this paper we have used natural unit where $c=1$.

Hence from eqs. (\ref{vacv}), (\ref{wj}), (\ref{wjl}), (\ref{ekcd}) and
(\ref{eeffl}) we find that the eikonal current for light-like charge generates pure gauge field in quantum field theory.

\section{ Proof of Factorization of $J/\Psi$ Production at High Energy Colliders }

Arguments about the proof of factorization of $j/\psi$ production in color singlet mechanism at high energy colliders
by using pQCD diagrammatic method in QCD in vacuum is given in \cite{singf}.
In this section we will prove the factorization theorem of $j/\psi$ production from color singlet and spin triplet $c{\bar c}$
pair at high energy colliders at all order in coupling constant
by using path integral formulation of QCD in vacuum. We will extend this path integral technique
to prove factorization theorem of $j/\psi$ production from color singlet and spin triplet $c{\bar c}$
pair in non-equilibrium QCD at RHIC and LHC at all order in coupling constant in section VII.

The generating functional in QCD including the heavy quark is given by \cite{muta,abbott}
\bea
&& Z[J,\eta_u,{\bar \eta}_u,\eta_d,{\bar \eta}_d,\eta_s,{\bar \eta}_s,\eta_h, {\bar \eta}_h]=\int [dQ] [d{\bar \psi}_1] [d \psi_1 ] [d{\bar \psi}_2] [d \psi_2 ][d{\bar \psi}_3] [d \psi_3 ][d{\bar \Psi}] [d \Psi ]~{\rm det}(\frac{\delta (\partial_\mu Q^{\mu a})}{\delta \omega^b}) \nonumber \\
&& e^{i\int d^4x [-\frac{1}{4}{F^a}_{\mu \nu}^2[Q] -\frac{1}{2 \alpha}
(\partial_\mu Q^{\mu a})^2+ J \cdot Q +\sum_{l=1}^3\left[{\bar \psi}_l  [i\gamma^\mu \partial_\mu -m_l +gT^a\gamma^\mu Q^a_\mu] \psi_l +{\bar \eta}_l \psi_l +  {\bar \psi}_l\eta_l\right] +{\bar \Psi}  [i\gamma^\mu \partial_\mu -M +gT^a\gamma^\mu Q^a_\mu] \Psi+{\bar \eta}_h \Psi + {\bar \Psi}\eta_h]}\nonumber \\
\label{aqcd}
\eea
where $Q^{\mu a}$ is the quantum gluon field, the symbols $l=1,2,3=u,d,s$ stand for three light quarks $u,d,s$ and the symbol $h$ stands for
heavy quark and
\bea
F_{\mu \nu}^a[Q]=\partial_\mu Q_\nu^a(x)-\partial_\nu Q_\mu^a(x)+gf^{abc} Q_\mu^b(x)Q_\nu^c(x),~~~~~~~~~~{F^a}_{\mu \nu}^2[Q]=F_{\mu \nu}^a[Q]F^{\mu \nu a}[Q].
\eea
Note that the determinant ${\rm det}(\frac{\delta (\partial_\mu Q^{\mu a})}{\delta \omega^b})$ in eq. (\ref{aqcd}) can be
expressed in terms of path integration over the ghost fields \cite{muta}. However, we will directly work with the
determinant ${\rm det}(\frac{\delta (\partial_\mu Q^{\mu a})}{\delta \omega^b})$ in eq. (\ref{aqcd}).

For the heavy quark Dirac field $\Psi(x)$,
the non-perturbative matrix element of the type $<0|{\bar \Psi}(x_1) \Psi(x_1) {\bar \Psi}(x_2) \Psi(x_2)|0>$ in
QCD is given by \cite{tucci}
\bea
&&<0|{\bar \Psi}(x_1) \Psi(x_1) {\bar \Psi}(x_2) \Psi(x_2)|0>\nonumber \\
&&=\int [dQ] [d{\bar \psi}_1] [d \psi_1 ] [d{\bar \psi}_2] [d \psi_2 ][d{\bar \psi}_3] [d \psi_3 ][d{\bar \Psi}] [d \Psi ]~{\bar \Psi}(x_1) \Psi(x_1) {\bar \Psi}(x_2) \Psi(x_2){\rm det}(\frac{\delta (\partial_\mu Q^{\mu a})}{\delta \omega^b}) \nonumber \\
&& e^{i\int d^4x [-\frac{1}{4}{F^a}_{\mu \nu}^2[Q] -\frac{1}{2 \alpha}
(\partial_\mu Q^{\mu a})^2+\sum_{l=1}^3\left[{\bar \psi}_l  [i\gamma^\mu \partial_\mu -m_l +gT^a\gamma^\mu Q^a_\mu] \psi_l \right] +{\bar \Psi}  [i\gamma^\mu \partial_\mu -M +gT^a\gamma^\mu Q^a_\mu] \Psi]}
\label{alepg}
\eea
where the suppression of the normalization factor $Z[0]$
is understood as it will cancel in the final result (see eq. (\ref{finalfacts})).

As mentioned in sections IV and V the infrared divergences can be studied by using eikonal approximation in quantum field theory.
We are interested in the infrared divergence behavior of the
non-perturbative heavy quark-antiquark correlation function of the type
$<0|{\bar \Psi}(x_1) \Psi(x_1) {\bar \Psi}(x_2) \Psi(x_2)|0>$ in the presence of light-like Wilson line in QCD.

The Eikonal propagator times the Eikonal vertex for a gluon with four momentum $k^\mu$ interacting with a
light-like quark moving with four velocity $l^\mu$ is given by
\cite{collins,tucci,collinssterman,berger,frederix,nayakqed,scet1,pathorder,nayaka2,nayaka3,nayaksterman,nayaksterman1}
\bea
gT^a~\frac{l^\mu}{l \cdot k+i\epsilon }.
\label{eikonalinp}
\eea
Hence for a light-like quark attached to infinite number of gluons we find the eikonal factor
\bea
&& 1+ gT^a\int \frac{d^4k}{(2\pi)^4} \frac{l\cdot { A}^a(k)}{l\cdot k +i\epsilon }
+g^2\int \frac{d^4k_1}{(2\pi)^4} \frac{d^4k_2}{(2\pi)^4} \frac{T^a l\cdot { A}^a(k_1)T^b l\cdot { A}^b(k_2)}{(l\cdot k_1 +i \epsilon)(l\cdot (k_1+k_2) +i \epsilon)}+... \nonumber \\
&&=1+ igT^a\int_0^{\infty} d\lambda l\cdot {  A}^a(l\lambda)+g^2i^2 \int_0^{\infty}  d\lambda_1 \int_{\lambda_1}^{\infty} d\lambda_2 T^a l\cdot { A}^a(l\lambda_1) T^b l\cdot { A}^b(l\lambda_2)+...\nonumber \\
&&= 1+ igT^a\int_0^{\infty} d\lambda l\cdot {  A}^a(l\lambda)+\frac{g^2i^2}{2!} {\cal P}\int_0^{\infty}  d\lambda_1 \int_0^{\infty}  d\lambda_2 T^a l\cdot { A}^a(l\lambda_1) T^b l\cdot { A}^b(l\lambda_2)+...
\nonumber \\
&&={\cal P}~{\rm exp}[ig \int_0^{\infty} d\lambda l\cdot { A}^a(l\lambda)T^a ]
\label{iiij}
\eea
which describes the infrared divergences arising from the infinite number of soft gluons exchange
with the light-like quark where ${\cal P}$ is  the path ordering and the gluon field $ { A}^{\mu a}(x)$ and its Fourier transform
$ { A}^{\mu a}(k)$ are related by
\bea
{ A}^{\mu a}(x) =\int \frac{d^4k}{(2\pi)^4} { A}^{\mu a}(k) e^{ik \cdot x}.
\label{ft}
\eea.
As mentioned in sections IV and V the light-like quark traveling with light-like
four-velocity $l^\mu$ produces SU(3) pure gauge field
at all the time-space position $x^\mu$ except at the position ${\vec x}$ perpendicular to the direction of motion
of the quark (${\vec l}\cdot {\vec x}=0$) at the time of closest approach \cite{collinssterman,nayakj,nayake}.
When $A^{\mu a}(x) = A^{\mu a}(\lambda l)$ as in eq. (\ref{iiij})
we find ${\vec l}\cdot {\vec x}=\lambda {\vec l}\cdot {\vec l}=\lambda\neq 0$ which implies that the light-like quark
finds the gluon field $A^{\mu a}(x)$ in eq. (\ref{iiij}) as the SU(3) pure gauge. The SU(3) pure gauge is given by
\bea
T^aA_\mu^a (x)= \frac{1}{ig}[\partial_\mu U(x)] ~U^{-1}(x),~~~~~~~~~~~~~U(x)=e^{igT^a\omega^a(x)}
\label{gtqcd}
\eea
which gives
\bea
U(x_f)={\cal P}e^{ig \int_{x_i}^{x_f} dx^\mu A_\mu^a(x) T^a}U(x_i)=e^{igT^a\omega^a(x_f)}.
\label{uxf}
\eea
Hence when $A^{\mu a}(x) = A^{\mu a}(\lambda l)$ as in eq. (\ref{iiij}) we find from eq.
(\ref{uxf}) that the light-like Wilson line in QCD for infrared divergences is given by
\cite{nayakall}
\bea
\Phi(x)={\cal P}e^{-ig \int_0^{\infty} d\lambda l\cdot { A}^a(x+l\lambda)T^a }=e^{igT^a\omega^a(x)}.
\label{ttt}
\eea
Hence the effect of infrared gluons interaction between the partons and the light-like Wilson line in QCD can be studied by
putting the partons in the SU(3) pure gauge background field which implies that the infrared behavior of the non-perturbative
correlation function of the type $<{\bar \Psi}(x) \Psi(x') {\bar \Psi}(x'') \Psi(x''')...>$ in QCD
due to the presence of light-like Wilson line in QCD can be studied by using the path integral method
of the QCD in the presence of SU(3) pure gauge background field as given by eq. (\ref{gtqcd}).

Background field method of QCD was originally formulated by 't Hooft \cite{thooft} and later
extended by Klueberg-Stern and Zuber \cite{zuber,zuber1} and by Abbott \cite{abbott}.
This is an elegant formalism which can be useful to construct gauge invariant
non-perturbative green's functions in QCD. This formalism is also useful to study quark and gluon production from classical chromo field
via Schwinger mechanism \cite{peter}, to compute $\beta$ function in QCD \cite{peskin}, to perform
calculations in lattice gauge theories \cite{lattice} and to study evolution of QCD
coupling constant in the presence of chromofield \cite{nayak}.

The generating functional in the background field method of QCD is given by
\bea
&& Z[A,J,\eta_u,{\bar \eta}_u,\eta_d,{\bar \eta}_d,\eta_s,{\bar \eta}_s,\eta_h, {\bar \eta}_h]=\int [dQ] [d{\bar \psi}_1] [d \psi_1 ] [d{\bar \psi}_2] [d \psi_2 ][d{\bar \psi}_3] [d \psi_3 ][d{\bar \Psi}] [d \Psi ]~{\rm det}(\frac{\delta G^a(Q)}{\delta \omega^b}) \nonumber \\
&& {\rm exp}[i\int d^4x [-\frac{1}{4}{F^a}_{\mu \nu}^2[A+Q] -\frac{1}{2 \alpha}
(G^a(Q))^2+ J \cdot Q \nonumber \\
&&+\sum_{l=1}^3\left[{\bar \psi}_l [i\gamma^\mu \partial_\mu -m_l +gT^a\gamma^\mu (A+Q)^a_\mu] \psi_l +{\bar \eta}_l \psi_l +  {\bar \psi}_l\eta_l\right] \nonumber \\
&&+{\bar \Psi} [i\gamma^\mu \partial_\mu -M +gT^a\gamma^\mu (A+Q)^a_\mu] \Psi+{\bar \eta}_h \Psi + {\bar \Psi}\eta_h]]
\label{azaqcd0}
\eea
where the gauge fixing term is given by
\bea
G^a(Q) =\partial_\mu Q^{\mu a} + gf^{abc} A_\mu^b Q^{\mu c}=D_\mu[A]Q^{\mu a}
\label{ga}
\eea
which depends on the background field $A^{\mu a}(x)$ and
\bea
F_{\mu \nu}^a[A+Q]=\partial_\mu [A_\nu^a+Q_\nu^a]-\partial_\nu [A_\mu^a+Q_\mu^a]+gf^{abc} [A_\mu^b+Q_\mu^b][A_\nu^c+Q_\nu^c].
\label{fqa}
\eea
We have followed the notations of \cite{thooft,zuber,abbott} and accordingly we have
denoted the quantum gluon field by $Q^{\mu a}$ and the background field by $A^{\mu a}$.
The gauge fixing term $\frac{1}{2 \alpha} (G^a(Q))^2$ in eq. (\ref{azaqcd0}) [where $G^a(Q)$ is given by eq. (\ref{ga})]
is invariant for gauge transformation of $A_\mu^a$:
\bea
\delta A_\mu^a = gf^{abc}A_\mu^b\omega^c + \partial_\mu \omega^a,  ~~~~~~~({\rm type~ I ~transformation})
\label{typeI}
\eea
provided one also performs a homogeneous transformation of $Q_\mu^a$ \cite{zuber,abbott}:
\bea
\delta Q_\mu^a =gf^{abc}Q_\mu^b\omega^c.
\label{omega}
\eea
The gauge transformation of background field $A_\mu^a$ as given by eq. (\ref{typeI})
along with the homogeneous transformation of $Q_\mu^a$ in eq. (\ref{omega}) gives
\bea
\delta (A_\mu^a+Q_\mu^a) = gf^{abc}(A_\mu^b+Q_\mu^b)\omega^c + \partial_\mu \omega^a
\label{omegavbxn}
\eea
which leaves $-\frac{1}{4}{F^a}_{\mu \nu}^2[A+Q]$ invariant in eq. (\ref{azaqcd0}).

For the heavy quark Dirac field $\Psi(x)$,
the non-perturbative matrix element of the type $<0|{\bar \Psi}(x_1) \Psi(x_1) {\bar \Psi}(x_2) \Psi(x_2)|0>$ in
the background field method of QCD is given by \cite{tucci}
\bea
&&<0|{\bar \Psi}(x_1) \Psi(x_1) {\bar \Psi}(x_2) \Psi(x_2)|0>_A\nonumber \\
&&=\int [dQ] [d{\bar \psi}_1] [d \psi_1 ] [d{\bar \psi}_2] [d \psi_2 ][d{\bar \psi}_3] [d \psi_3 ][d{\bar \Psi}] [d \Psi ]~{\bar \Psi}(x_1) \Psi(x_1) {\bar \Psi}(x_2) \Psi(x_2)
~{\rm det}(\frac{\delta G^a(Q)}{\delta \omega^b}) \nonumber \\
&& {\rm exp}[i\int d^4x [-\frac{1}{4}{F^a}_{\mu \nu}^2[A+Q] -\frac{1}{2 \alpha}
(G^a(Q))^2 +\sum_{l=1}^3\left[{\bar \psi}_l [i\gamma^\mu \partial_\mu -m_l +gT^a\gamma^\mu (A+Q)^a_\mu] \psi_l \right] \nonumber \\
&&+{\bar \Psi} [i\gamma^\mu \partial_\mu -M +gT^a\gamma^\mu (A+Q)^a_\mu] \Psi]]
\label{balepg}
\eea
where the suppression of the normalization factor $Z[0]$
is understood as it will cancel in the final result (see eq. (\ref{finalfacts})).
By changing $Q \rightarrow Q-A$ in eq. (\ref{balepg}) we find that
\bea
&&<0|{\bar \Psi}(x_1) \Psi(x_1) {\bar \Psi}(x_2) \Psi(x_2)|0>_A\nonumber \\
&&=\int [dQ] [d{\bar \psi}_1] [d \psi_1 ] [d{\bar \psi}_2] [d \psi_2 ][d{\bar \psi}_3] [d \psi_3 ][d{\bar \Psi}] [d \Psi ]~{\bar \Psi}(x_1) \Psi(x_1) {\bar \Psi}(x_2) \Psi(x_2)
~{\rm det}(\frac{\delta G^a_f(Q)}{\delta \omega^b}) \nonumber \\
&& {\rm exp}[i\int d^4x [-\frac{1}{4}{F^a}_{\mu \nu}^2[Q] -\frac{1}{2 \alpha}
(G^a_f(Q))^2 +\sum_{l=1}^3\left[{\bar \psi}_l [i\gamma^\mu \partial_\mu -m_l +gT^a\gamma^\mu Q^a_\mu] \psi_l \right] \nonumber \\
&&+{\bar \Psi} [i\gamma^\mu \partial_\mu -M +gT^a\gamma^\mu Q^a_\mu] \Psi]]
\label{zaqcd1}
\eea
where the gauge fixing term from eq. (\ref{ga}) becomes
\bea
G_f^a(Q) =\partial_\mu Q^{\mu a} + gf^{abc} A_\mu^b Q^{\mu c} - \partial_\mu A^{\mu a}=D_\mu[A] Q^{\mu a} - \partial_\mu A^{\mu a}
\label{gfa}
\eea
and from eq. (\ref{omega}) [by using eq. (\ref{typeI})] we find
\bea
\delta Q_\mu^a = -gf^{abc}\omega^b Q_\mu^c + \partial_\mu \omega^a.
\label{theta}
\eea
For finite transformation eq. (\ref{theta}) becomes
\bea
T^aQ'^a_\mu(x) = U(x)T^aQ^a_\mu(x) U^{-1}(x)+\frac{1}{ig}[\partial_\mu U(x)] U^{-1}(x),~~~~~~~~~~~U(x)=e^{igT^a\omega^a(x)}.
\label{te}
\eea
The fermion fields transform as
\bea
\psi'_l(x)=e^{igT^a\omega^a(x)}\psi_l(x),~~~~~~~~~~~~~\Psi'(x)=e^{igT^a\omega^a(x)}\Psi(x).
\label{pg3}
\eea
Changing the variables of integration from unprimed to primed variables in eq. (\ref{zaqcd1}) we find
\bea
&&<0|{\bar \Psi}(x_1) \Psi(x_1) {\bar \Psi}(x_2) \Psi(x_2)|0>_A\nonumber \\
&&=\int [dQ'] [d{\bar \psi}'_1] [d \psi'_1 ] [d{\bar \psi}'_2] [d \psi'_2 ][d{\bar \psi}'_3] [d \psi'_3 ][d{\bar \Psi}'] [d \Psi' ]~{\bar \Psi}'(x_1) \Psi'(x_1) {\bar \Psi}'(x_2) \Psi'(x_2)
~{\rm det}(\frac{\delta G^a_f(Q')}{\delta \omega^b}) \nonumber \\
&& {\rm exp}[i\int d^4x [-\frac{1}{4}{F^a}_{\mu \nu}^2[Q'] -\frac{1}{2 \alpha}
(G^a_f(Q'))^2 +\sum_{l=1}^3\left[{\bar \psi}'_l [i\gamma^\mu \partial_\mu -m_l +gT^a\gamma^\mu Q'^a_\mu] \psi'_l \right] \nonumber \\
&&+{\bar \Psi}' [i\gamma^\mu \partial_\mu -M +gT^a\gamma^\mu Q'^a_\mu] \Psi']]
\label{zaqcd1b}
\eea
because the change of variables from unprimed to primed variables does not change the value of the integration.

From eqs. (\ref{te}) and (\ref{pg3}) we find \cite{nayakall}
\bea
&& [dQ'] =[dQ],~~~~~[d{\bar \psi}_1'] [d \psi'_1 ]=[d{\bar \psi}_1] [d \psi_1 ],~~~~~~~[d{\bar \psi}'_2] [d \psi'_2 ]=[d{\bar \psi}_2] [d \psi_2 ],~~~~~~[d{\bar \psi}'_3] [d \psi'_3 ]=[d{\bar \psi}_3] [d \psi_3 ],\nonumber \\
&&[d{\bar \Psi}'] [d \Psi' ]=[d{\bar \Psi}] [d \Psi ],~~~~~~{\bar \psi}'_l [i\gamma^\mu \partial_\mu -m_l +gT^a\gamma^\mu Q'^a_\mu] \psi'_l={\bar \psi}_l [i\gamma^\mu \partial_\mu -m_l +gT^a\gamma^\mu Q^a_\mu] \psi_l, \nonumber \\
&&{\bar \Psi}' [i\gamma^\mu \partial_\mu -M +gT^a\gamma^\mu Q'^a_\mu] \Psi'={\bar \Psi} [i\gamma^\mu \partial_\mu -M +gT^a\gamma^\mu Q^a_\mu]\Psi,~~~~~~~~~{F^a}_{\mu \nu}^2[Q']={F^a}_{\mu \nu}^2[Q] \nonumber \\
&& (G_f^a(Q'))^2 = (\partial_\mu Q^{\mu a}(x))^2,~~~~~~~~~~~{\rm det} [\frac{\delta G_f^a(Q')}{\delta \omega^b}] ={\rm det}[\frac{ \delta (\partial_\mu Q^{\mu a}(x))}{\delta \omega^b}].
\label{gqp4a}
\eea
Using eq. (\ref{gqp4a}) in (\ref{zaqcd1b}) we find
\bea
&&<0|{\bar \Psi}(x_1) \Psi(x_1) {\bar \Psi}(x_2) \Psi(x_2)|0>\nonumber \\
&&=\int [dQ] [d{\bar \psi}_1] [d \psi_1 ] [d{\bar \psi}_2] [d \psi_2 ][d{\bar \psi}_3] [d \psi_3 ][d{\bar \Psi}] [d \Psi ]~{\bar \Psi}(x_1) \Psi(x_1) {\bar \Psi}(x_2) \Psi(x_2){\rm det}(\frac{\delta (\partial_\mu Q^{\mu a})}{\delta \omega^b}) \nonumber \\
&& e^{i\int d^4x [-\frac{1}{4}{F^a}_{\mu \nu}^2[Q] -\frac{1}{2 \alpha}
(\partial_\mu Q^{\mu a})^2+\sum_{l=1}^3\left[{\bar \psi}_l  [i\gamma^\mu \partial_\mu -m_l +gT^a\gamma^\mu Q^a_\mu] \psi_l \right] +{\bar \Psi}  [i\gamma^\mu \partial_\mu -M +gT^a\gamma^\mu Q^a_\mu] \Psi]}.
\label{falepg}
\eea
From eqs. (\ref{falepg}) and (\ref{alepg}) we find
\bea
&&<0|{\bar \Psi}(x_1) \Psi(x_1) {\bar \Psi}(x_2) \Psi(x_2)|0>=<0|{\bar \Psi}(x_1) \Psi(x_1) {\bar \Psi}(x_2)\Psi(x_2)|0>_A
\label{finalfacts}
\eea
which proves factorization of infrared divergences at all order in coupling constant.
Eq. (\ref{finalfacts}) is valid in covariant gauge, in light-cone gauge, in general axial gauges, in general non-covariant
gauges and in general Coulomb gauge etc. respectively \cite{nayakall}.

From eq. (\ref{finalfacts}) we find that the non-perturbative matrix element
for $j/\psi$ production from $c{\bar c}$ pair in color singlet and spin triplet state in eq. (\ref{alll}) as given by
\bea
<0|{\cal O}_{H}|0> =<0|\chi^\dagger \sigma^i \psi a^\dagger_Ha_H \psi^\dagger \sigma^i \chi |0>
\label{alllf}
\eea
is consistent with factorization theorem of $j/\psi$ production at high energy colliders
at all order in coupling constant.

Eq. (\ref{finalfacts}) proves the factorization theorem of $j/\psi$ production at high energy colliders at all order in coupling constant
where $<0|{\bar \Psi}(x_1) \Psi(x_1) {\bar \Psi}(x_2) \Psi(x_2)|0>$
is the heavy quark-antiquark gauge invariant non-perturbative correlation function in QCD and
$<0|{\bar \Psi}(x_1) \Psi(x_1) {\bar \Psi}(x_2) \Psi(x_2)|0>_A$ is the corresponding
heavy quark-antiquark gauge invariant non-perturbative correlation function in QCD in the presence of light-like
Wilson line. From eq. (\ref{finalfacts}) one finds that the infrared divergences due to the soft gluons
exchange between charm quark and the nearby light-like quark (or gluon) at high energy colliders
exactly cancel with the corresponding infrared divergences
due to the soft gluons exchange between anticharm quark and the same nearby light-like quark (or gluon)
at all order in coupling constant in the
$j/\psi$ production from color singlet $c{\bar c}$ pair at high energy colliders.
This proves the factorization theorem of $j/\psi$ production at high energy colliders
at all order in coupling constant.

\section{ Proof of factorization of $J/\Psi$ production in non-equilibrium QCD at RHIC and LHC }

The generating functional in non-equilibrium QCD (including heavy quark) in the path integral formulation is given by eq.
(\ref{zfqinon}). The nonequilibrium-nonperturbative heavy quark-antiquark correlation function of the type
$<in|{\bar \Psi}_r(x_1) \Psi_r(x_1) {\bar \Psi}_s(x_2) \Psi_s(x_2)|in>$ in QCD is given by eq. (\ref{malepg}) which gives
\bea
&& <in|{\bar \Psi}_r(x_1) \Psi_r(x_1) {\bar \Psi}_s(x_2) \Psi_s(x_2)|in>=\nonumber \\
&&=\int [dQ_+] [dQ_-][d{\bar \psi}_{1+}] [d{\bar \psi}_{1-}] [d \psi_{1+} ] [d\psi_{1-}][d{\bar \psi}_{2+}] [d{\bar \psi}_{2-}] [d \psi_{2+} ] [d\psi_{2-}][d{\bar \psi}_{3+}] [d{\bar \psi}_{3-}] [d \psi_{3+} ] [d\psi_{3-}]\nonumber \\
&&
[d{\bar \Psi}_{+}] [d{\bar \Psi}_{-}] [d \Psi_{+} ] [d\Psi_{-}]~{\bar \Psi}_r(x_1) \Psi_r(x_1) {\bar \Psi}_s(x_2) \Psi_s(x_2)
\times ~{\rm det}(\frac{\delta \partial_\mu Q_+^{\mu a}}{\delta \omega_+^b})~\times ~{\rm det}(\frac{\delta \partial_\mu Q_-^{\mu a}}{\delta \omega_-^b}) \nonumber \\
&& {\rm exp}[i\int d^4x [-\frac{1}{4}({F^a}_{\mu \nu}^2[Q_+]-{F^a}_{\mu \nu}^2[Q_-])-\frac{1}{2 \alpha} (
(\partial_\mu Q_+^{\mu a })^2-(\partial_\mu Q_-^{\mu a })^2) \nonumber \\
&&+\sum_{l=1}^3{\bar \psi}_{l+}  [i\gamma^\mu \partial_\mu -m_l +gT^a\gamma^\mu Q^a_{\mu +}]  \psi_{l+} -\sum_{l=1}^3{\bar \psi}_{l-}  [i\gamma^\mu \partial_\mu -m_l +gT^a\gamma^\mu Q^a_{\mu -}]  \psi_{l-}\nonumber \\
&&+{\bar \Psi}_{+}  [i\gamma^\mu \partial_\mu -M +gT^a\gamma^\mu Q^a_{\mu +}]  \Psi_{+}-{\bar \Psi}_{-}  [i\gamma^\mu \partial_\mu -M +gT^a\gamma^\mu Q^a_{\mu -}]  \Psi_{-}]]\nonumber \\
&& \times ~<Q_+,\psi_+^u,{\bar \psi}_+^u,\psi_+^d,{\bar \psi}_+^d,\psi_+^s,{\bar \psi}_+^s,\Psi_+,{\bar \Psi}_+,0|~\rho~|0,\Psi_-,{\bar \Psi}_-,{\bar \psi}_-^s,\psi_-^s,{\bar \psi}_-^d,\psi_-^d,{\bar \psi}_-^u,\psi_-^u,Q_-> \nonumber \\
\label{nezf}
\eea
where the repeated (closed-time path) indices $r,s=+,-$ are not summed and the suppression of the normalization factor
$Z[0]$ is understood as it will cancel in the final result, see eq. (\ref{finalz}).

The generating functional in the background field method of QCD in
non-equilibrium QCD (including heavy quark) in the path integral formulation is given by \cite{greiner,cooper,thooft,abbott,zuber,zuber1}
\bea
&& Z[\rho,A,J_+,J_-,\eta_+^u,\eta_-^u,{\bar \eta}_+^u,{\bar \eta}_-^u,\eta_+^d,\eta_-^d,{\bar \eta}_+^d,{\bar \eta}_-^d,\eta_+^s,\eta_-^s,{\bar \eta}_+^s,{\bar \eta}_-^s,\eta_+^h,\eta_-^h,{\bar \eta}_+^h,{\bar \eta}_-^h]\nonumber \\
&&=\int [dQ_+] [dQ_-][d{\bar \psi}_{1+}] [d{\bar \psi}_{1-}] [d \psi_{1+} ] [d\psi_{1-}][d{\bar \psi}_{2+}] [d{\bar \psi}_{2-}] [d \psi_{2+} ] [d\psi_{2-}][d{\bar \psi}_{3+}] [d{\bar \psi}_{3-}] [d \psi_{3+} ] [d\psi_{3-}]\nonumber \\
&&
[d{\bar \Psi}_{+}] [d{\bar \Psi}_{-}] [d \Psi_{+} ] [d\Psi_{-}] \times {\rm det}(\frac{\delta G^a(Q_+)}{\delta \omega_+^b})~\times ~{\rm det}(\frac{\delta G^a(Q_-)}{\delta \omega_-^b}) \times {\rm exp}[i\int d^4x [-\frac{1}{4}({F^a}_{\mu \nu}^2[Q_++A_+]\nonumber \\
&&-{F^a}_{\mu \nu}^2[Q_-+A_-])-\frac{1}{2 \alpha} ((G^a(Q_+))^2-(G^a(Q_-))^2)
+\sum_{l=1}^3{\bar \psi}_{l+}  [i\gamma^\mu \partial_\mu -m_l +gT^a\gamma^\mu (Q_++A_+)^a_\mu]  \psi_{l+} \nonumber \\
&&-\sum_{l=1}^3{\bar \psi}_{l-}  [i\gamma^\mu \partial_\mu -m_l +gT^a\gamma^\mu (Q_-+A_-)^a_\mu]  \psi_{l-}+{\bar \Psi}_{+}  [i\gamma^\mu \partial_\mu -M +gT^a\gamma^\mu (Q_++A_+)^a_\mu]  \Psi_{+}\nonumber \\
&&-{\bar \Psi}_{-}  [i\gamma^\mu \partial_\mu -M +gT^a\gamma^\mu (Q_-+A_-)^a_\mu]  \Psi_{-}
+ J_+ \cdot Q_+ -J_- \cdot Q_-\nonumber \\
&&+\sum_{l=1}^3[{\bar \eta}_{l+} \cdot \psi_{l+} - {\bar \eta}_{l-} \cdot \psi_{l-} +  {\bar \psi}_{l+} \cdot \eta_{l+}
- {\bar \psi}_{l-} \cdot \eta_{l-}]+{\bar \eta}_{h+} \cdot \Psi_{+} - {\bar \eta}_{h-} \cdot \Psi_{-} +  {\bar \Psi}_{+} \cdot \eta_{h+}
- {\bar \Psi}_{-} \cdot \eta_{h-}]]
\nonumber \\
&& \times ~<Q_++A_+,\psi_+^u,{\bar \psi}_+^u,\psi_+^d,{\bar \psi}_+^d,\psi_+^s,{\bar \psi}_+^s,\Psi_+,{\bar \Psi}_+,0|~\rho~|0,\Psi_-,{\bar \Psi}_-,{\bar \psi}_-^s,\psi_-^s,{\bar \psi}_-^d,\psi_-^d,{\bar \psi}_-^u,\psi_-^u,Q_-+A_->\nonumber \\
\label{azaqcd}
\eea
where the gauge fixing term $G^a(Q)$ is given by eq. (\ref{ga}) which depends on the background field $A^{\mu a}(x)$, the
$F_{\mu \nu}^a[Q+A]$ is given by eq. (\ref{fqa}), the $\delta Q_\mu^a$ and $\delta A_\mu^a$ are given by eqs. (\ref{omega})
and (\ref{typeI}) respectively.

The nonequilibrium-nonperturbative heavy quark-antiquark correlation function of the type
$<in|{\bar \Psi}_r(x_1) \Psi_r(x_1) {\bar \Psi}_s(x_2) \Psi_s(x_2)|in>$ in the background field
method of QCD is given by
\bea
&& <in|{\bar \Psi}_r(x_1) \Psi_r(x_1) {\bar \Psi}_s(x_2) \Psi_s(x_2)|in>_A\nonumber \\
&&=\int [dQ_+] [dQ_-][d{\bar \psi}_{1+}] [d{\bar \psi}_{1-}] [d \psi_{1+} ] [d\psi_{1-}][d{\bar \psi}_{2+}] [d{\bar \psi}_{2-}] [d \psi_{2+} ] [d\psi_{2-}][d{\bar \psi}_{3+}] [d{\bar \psi}_{3-}] [d \psi_{3+} ] [d\psi_{3-}]\nonumber \\
&&
[d{\bar \Psi}_{+}] [d{\bar \Psi}_{-}] [d \Psi_{+} ] [d\Psi_{-}]~{\bar \Psi}_r(x_1) \Psi_r(x_1) {\bar \Psi}_s(x_2) \Psi_s(x_2)\nonumber \\
&&
\times {\rm det}(\frac{\delta G^a(Q_+)}{\delta \omega_+^b})~\times ~{\rm det}(\frac{\delta G^a(Q_-)}{\delta \omega_-^b}) \times {\rm exp}[i\int d^4x [-\frac{1}{4}({F^a}_{\mu \nu}^2[Q_++A_+]\nonumber \\
&&-{F^a}_{\mu \nu}^2[Q_-+A_-])-\frac{1}{2 \alpha} ((G^a(Q_+))^2-(G^a(Q_-))^2)
+\sum_{l=1}^3{\bar \psi}_{l+}  [i\gamma^\mu \partial_\mu -m_l +gT^a\gamma^\mu (Q_++A_+)^a_\mu]  \psi_{l+} \nonumber \\
&&-\sum_{l=1}^3{\bar \psi}_{l-}  [i\gamma^\mu \partial_\mu -m_l +gT^a\gamma^\mu (Q_-+A_-)^a_\mu]  \psi_{l-}+{\bar \Psi}_{+}  [i\gamma^\mu \partial_\mu -M +gT^a\gamma^\mu (Q_++A_+)^a_\mu]  \Psi_{+}\nonumber \\
&&-{\bar \Psi}_{-}  [i\gamma^\mu \partial_\mu -M +gT^a\gamma^\mu (Q_-+A_-)^a_\mu]  \Psi_{-}
]]
\nonumber \\
&& \times ~<Q_++A_+,\psi_+^u,{\bar \psi}_+^u,\psi_+^d,{\bar \psi}_+^d,\psi_+^s,{\bar \psi}_+^s,\Psi_+,{\bar \Psi}_+,0|~\rho~|0,\Psi_-,{\bar \Psi}_-,{\bar \psi}_-^s,\psi_-^s,{\bar \psi}_-^d,\psi_-^d,{\bar \psi}_-^u,\psi_-^u,Q_-+A_->\nonumber \\
\label{cfqcd}
\eea
where the suppression of the normalization factor $Z[0]$ is understood as it will cancel in the final result, see eq. (\ref{finalz}).

In order to study the infrared behavior of the nonequilibrium-nonperturbative heavy quark-antiquark correlation function
we proceed as follows. By changing the integration variable $Q \rightarrow Q-A$ in the right hand side of eq. (\ref{cfqcd}) we find
\bea
&& <in|{\bar \Psi}_r(x_1) \Psi_r(x_1) {\bar \Psi}_s(x_2) \Psi_s(x_2)|in>_A\nonumber \\
&&=\int [dQ_+] [dQ_-][d{\bar \psi}_{1+}] [d{\bar \psi}_{1-}] [d \psi_{1+} ] [d\psi_{1-}][d{\bar \psi}_{2+}] [d{\bar \psi}_{2-}] [d \psi_{2+} ] [d\psi_{2-}][d{\bar \psi}_{3+}] [d{\bar \psi}_{3-}] [d \psi_{3+} ] [d\psi_{3-}]\nonumber \\
&&
[d{\bar \Psi}_{+}] [d{\bar \Psi}_{-}] [d \Psi_{+} ] [d\Psi_{-}]~{\bar \Psi}_r(x_1) \Psi_r(x_1) {\bar \Psi}_s(x_2) \Psi_s(x_2)\nonumber \\
&&
\times {\rm det}(\frac{\delta G^a_f(Q_+)}{\delta \omega_+^b})~\times ~{\rm det}(\frac{\delta G^a_f(Q_-)}{\delta \omega_-^b}) \times {\rm exp}[i\int d^4x [-\frac{1}{4}({F^a}_{\mu \nu}^2[Q_+]\nonumber \\
&&-{F^a}_{\mu \nu}^2[Q_-])-\frac{1}{2 \alpha} ((G^a_f(Q_+))^2-(G^a_f(Q_-))^2)
+\sum_{l=1}^3{\bar \psi}_{l+}  [i\gamma^\mu \partial_\mu -m_l +gT^a\gamma^\mu Q^a_{\mu +}]  \psi_{l+} \nonumber \\
&&-\sum_{l=1}^3{\bar \psi}_{l-}  [i\gamma^\mu \partial_\mu -m_l +gT^a\gamma^\mu Q^a_{\mu -}]  \psi_{l-}+{\bar \Psi}_{+}  [i\gamma^\mu \partial_\mu -M +gT^a\gamma^\mu Q^a_{\mu +}]  \Psi_{+}\nonumber \\
&&-{\bar \Psi}_{-}  [i\gamma^\mu \partial_\mu -M +gT^a\gamma^\mu Q^a_{\mu -}]  \Psi_{-}
]]
\nonumber \\
&& \times ~<Q_+,\psi_+^u,{\bar \psi}_+^u,\psi_+^d,{\bar \psi}_+^d,\psi_+^s,{\bar \psi}_+^s,\Psi_+,{\bar \Psi}_+,0|~\rho~|0,\Psi_-,{\bar \Psi}_-,{\bar \psi}_-^s,\psi_-^s,{\bar \psi}_-^d,\psi_-^d,{\bar \psi}_-^u,\psi_-^u,Q_->\nonumber \\
\label{cfqcd1}
\eea
where $G_f^a(Q)$ is given by eq. (\ref{gfa}) and eq. (\ref{omega}) [by using eq. (\ref{typeI})] becomes
eq. (\ref{theta}) which for finite transformation gives eq. (\ref{te}).
Changing the integration variable from unprimed variable to primed variable we find from eq. (\ref{cfqcd1})
\bea
&& <in|{\bar \Psi}_r(x_1) \Psi_r(x_1) {\bar \Psi}_s(x_2) \Psi_s(x_2)|in>_A\nonumber \\
&&=\int [dQ'_+] [dQ'_-][d{\bar \psi}'_{1+}] [d{\bar \psi}'_{1-}] [d \psi'_{1+} ] [d\psi'_{1-}][d{\bar \psi}'_{2+}] [d{\bar \psi}'_{2-}] [d \psi'_{2+} ] [d\psi'_{2-}][d{\bar \psi}'_{3+}] [d{\bar \psi}'_{3-}] [d \psi'_{3+} ] [d\psi'_{3-}]\nonumber \\
&&
[d{\bar \Psi}'_{+}] [d{\bar \Psi}'_{-}] [d \Psi'_{+} ] [d\Psi'_{-}]~{\bar \Psi}'_r(x_1) \Psi'_r(x_1) {\bar \Psi}'_s(x_2) \Psi'_s(x_2)\nonumber \\
&&
\times {\rm det}(\frac{\delta G^a_f(Q'_+)}{\delta \omega_+^b})~\times ~{\rm det}(\frac{\delta G^a_f(Q'_-)}{\delta \omega_-^b}) \times {\rm exp}[i\int d^4x [-\frac{1}{4}({F^a}_{\mu \nu}^2[Q'_+]-{F^a}_{\mu \nu}^2[Q'_-])-\frac{1}{2 \alpha} ((G^a_f(Q'_+))^2\nonumber \\
&&-(G^a_f(Q'_-))^2)
+\sum_{l=1}^3{\bar \psi}'_{l+}  [i\gamma^\mu \partial_\mu -m_l +gT^a\gamma^\mu Q'^a_{\mu +}]  \psi'_{l+} -\sum_{l=1}^3{\bar \psi}'_{l-}  [i\gamma^\mu \partial_\mu -m_l +gT^a\gamma^\mu Q'^a_{\mu -}]  \psi'_{l-}\nonumber \\
&&+{\bar \Psi}'_{+}  [i\gamma^\mu \partial_\mu -M +gT^a\gamma^\mu Q'^a_{\mu +}]  \Psi'_{+}-{\bar \Psi}'_{-}  [i\gamma^\mu \partial_\mu -M +gT^a\gamma^\mu Q'^a_{\mu -}]  \Psi'_{-}
]]
\nonumber \\
&& \times ~<Q'_+,\psi'^u_+,{\bar \psi}'^u_+,\psi'^d_+,{\bar \psi}'^d_+,\psi'^s_+,{\bar \psi}'^s_+,\Psi'_+,{\bar \Psi}'_+,0|~\rho~|0,\Psi'_-,{\bar \Psi}'_-,{\bar \psi}'^s_-,\psi'^s_-,{\bar \psi}'^d_-,\psi'^d_-,{\bar \psi}'^u_-,\psi'^u_-,Q'_->.\nonumber \\
\label{cfqcd1vb}
\eea
This is because a change of integration variable from unprimed variable to primed variable does not change the value of the
integration. Note that since we are working in the frozen ghost formalism at the initial time \cite{greiner,cooper} the
$<Q_+,\psi_+^u,{\bar \psi}_+^u,\psi_+^d,{\bar \psi}_+^d,\psi_+^s,{\bar \psi}_+^s,\Psi_+,{\bar \Psi}_+,0|~\rho~|0,\Psi_-,{\bar \Psi}_-,{\bar \psi}_-^s,\psi_-^s,{\bar \psi}_-^d,\psi_-^d,{\bar \psi}_-^u,\psi_-^u,Q_->$ in eq. (\ref{zfqinon}) corresponding to initial density of
state in non-equilibrium QCD is gauge invariant by definition. Hence from eqs. (\ref{te}) and (\ref{pg3}) we find
\bea
&&<Q'_+,\psi'^u_+,{\bar \psi}'^u_+,\psi'^d_+,{\bar \psi}'^d_+,\psi'^s_+,{\bar \psi}'^s_+,\Psi'_+,{\bar \Psi}'_+,0|~\rho~|0,\Psi'_-,{\bar \Psi}'_-,{\bar \psi}'^s_-,\psi'^s_-,{\bar \psi}'^d_-,\psi'^d_-,{\bar \psi}'^u_-,\psi'^u_-,Q'_-> \nonumber \\
&&=<Q_+,\psi_+^u,{\bar \psi}_+^u,\psi_+^d,{\bar \psi}_+^d,\psi_+^s,{\bar \psi}_+^s,\Psi_+,{\bar \Psi}_+,0|~\rho~|0,\Psi_-,{\bar \Psi}_-,{\bar \psi}_-^s,\psi_-^s,{\bar \psi}_-^d,\psi_-^d,{\bar \psi}_-^u,\psi_-^u,Q_->.\nonumber \\
\label{nmm}
\eea
We follow the similar procedure that was employed to prove factorization theorem of $j/\psi$ production in QCD in vacuum
in the previous section. When background field $A^{\mu a}(x)$ is the SU(3) pure gauge as given by eq. (\ref{gtqcd}) then by
using eqs. (\ref{gqp4a}) and (\ref{nmm}) in eq. (\ref{cfqcd1vb}) we find
\bea
&& <in|{\bar \Psi}_r(x_1) \Psi_r(x_1) {\bar \Psi}_s(x_2) \Psi_s(x_2)|in>_A=\nonumber \\
&&=\int [dQ_+] [dQ_-][d{\bar \psi}_{1+}] [d{\bar \psi}_{1-}] [d \psi_{1+} ] [d\psi_{1-}][d{\bar \psi}_{2+}] [d{\bar \psi}_{2-}] [d \psi_{2+} ] [d\psi_{2-}][d{\bar \psi}_{3+}] [d{\bar \psi}_{3-}] [d \psi_{3+} ] [d\psi_{3-}]\nonumber \\
&&
[d{\bar \Psi}_{+}] [d{\bar \Psi}_{-}] [d \Psi_{+} ] [d\Psi_{-}]~{\bar \Psi}_r(x_1) \Psi_r(x_1) {\bar \Psi}_s(x_2) \Psi_s(x_2) \times ~{\rm det}(\frac{\delta \partial_\mu Q_+^{\mu a}}{\delta \omega_+^b})~\times ~{\rm det}(\frac{\delta \partial_\mu Q_-^{\mu a}}{\delta \omega_-^b}) \nonumber \\
&& {\rm exp}[i\int d^4x [-\frac{1}{4}({F^a}_{\mu \nu}^2[Q_+]-{F^a}_{\mu \nu}^2[Q_-])-\frac{1}{2 \alpha} (
(\partial_\mu Q_+^{\mu a })^2-(\partial_\mu Q_-^{\mu a })^2) \nonumber \\
&&+\sum_{l=1}^3{\bar \psi}_{l+}  [i\gamma^\mu \partial_\mu -m_l +gT^a\gamma^\mu Q^a_{\mu +}]  \psi_{l+} -\sum_{l=1}^3{\bar \psi}_{l-}  [i\gamma^\mu \partial_\mu -m_l +gT^a\gamma^\mu Q^a_{\mu -}]  \psi_{l-}\nonumber \\
&&+{\bar \Psi}_{+}  [i\gamma^\mu \partial_\mu -M +gT^a\gamma^\mu Q^a_{\mu +}]  \Psi_{+}-{\bar \Psi}_{-}  [i\gamma^\mu \partial_\mu -M +gT^a\gamma^\mu Q^a_{\mu -}]  \Psi_{-}]]\nonumber \\
&& \times ~<Q_+,\psi_+^u,{\bar \psi}_+^u,\psi_+^d,{\bar \psi}_+^d,\psi_+^s,{\bar \psi}_+^s,\Psi_+,{\bar \Psi}_+,0|~\rho~|0,\Psi_-,{\bar \Psi}_-,{\bar \psi}_-^s,\psi_-^s,{\bar \psi}_-^d,\psi_-^d,{\bar \psi}_-^u,\psi_-^u,Q_->. \nonumber \\
\label{cfq5p2}
\eea
From eqs. (\ref{nezf}) and (\ref{cfq5p2}) we find
\bea
<in|{\bar \Psi}_r(x_1) \Psi_r(x_1) {\bar \Psi}_s(x_2) \Psi_s(x_2)|in>=<in|{\bar \Psi}_r(x_1) \Psi_r(x_1) {\bar \Psi}_s(x_2) \Psi_s(x_2|in>_A
\label{finalz}
\eea
which proves factorization of infrared divergences at all order in coupling constant in non-equilibrium QCD.
Eq. (\ref{finalz}) is valid in covariant gauge, in light-cone gauge, in general axial gauges, in general non-covariant
gauges and in general Coulomb gauge etc. respectively \cite{nayakall}.

From eq. (\ref{finalz}) we find that the nonequilibrium-nonperturbative matrix element
for $j/\psi$ production from $c{\bar c}$ pair in color singlet and spin triplet state in non-equilibrium QCD
in eq. (\ref{malll}) as given by
\bea
<in|{\cal O}_{H}|in> =<in|\chi^\dagger \sigma^i \psi a^\dagger_Ha_H \psi^\dagger \sigma^i \chi |in>
\label{malllf}
\eea
is consistent with factorization theorem of $j/\psi$ production in non-equilibrium QCD at RHIC and LHC at all order in coupling constant.

Eq. (\ref{finalz}) proves the factorization theorem of $j/\psi$ production in non-equilibrium QCD at RHIC and LHC
at all order in coupling constant where $<in|{\bar \Psi}_r(x_1) \Psi_r(x_1) {\bar \Psi}_s(x_2) \Psi_s(x_2)|in>$
is the heavy quark-antiquark gauge invariant nonequilibrium-nonperturbative correlation function in QCD and
$<in|{\bar \Psi}_r(x_1) \Psi_r(x_1) {\bar \Psi}_s(x_2) \Psi_s(x_2)|in>_A$ is the corresponding
heavy quark-antiquark gauge invariant nonequilibrium-nonperturbative correlation function in QCD in the presence of light-like
Wilson line. From eq. (\ref{finalz}) we find that the infrared divergences due to the soft gluons
exchange between charm quark and the nearby light-like quark (or gluon) in non-equilibrium
QCD at RHIC and LHC exactly cancel with the corresponding infrared divergences
due to the soft gluons exchange between anticharm quark and the same nearby light-like quark (or gluon)
at all order in coupling constant in the
$j/\psi$ production from color singlet $c{\bar c}$ pair in non-equilibrium QCD.
This proves the factorization theorem of $j/\psi$ production in non-equilibrium
QCD at RHIC and LHC at all order in coupling constant.

\section{Conclusions}

$J/\psi$ suppression/production is one of the main signature of quark-gluon plasma detection at RHIC and
LHC. In order to study $j/\psi$ suppression/production in high energy heavy-ion collisions at RHIC and LHC,
one needs to prove the factorization theorem of $j/\psi$ production in non-equilibrium QCD medium,
otherwise one will predict infinite cross section of $j/\psi$.
In this paper we have proved factorization theorem of $j/\psi$ production in non-equilibrium QCD at RHIC and LHC
at all order in coupling constant.
This proof is necessary to study the detection of quark-gluon plasma \cite{qgp,qgp1,qgp2,qgp3} at RHIC
and LHC by using $j/\psi$ suppression/production as its signature \cite{satz}.

\end{document}